\documentclass[12pt]{article}

\usepackage{graphicx}

\begin{document}

\title{About an alternative distribution function for fractional exclusion statistics}

\author{Qiuping A. Wang, A. Le M\'ehaut\'e, L. Nivanen \\ Institut Sup\'erieur des Mat\'eriaux du Mans, \\
44, Avenue F.A. Bartholdi, 72000 Le Mans, France \\ and M. Pezeril \\ Laboratoire de Physique
de l'\'etat Condens\'e, \\ Universit\'e du Maine, 72000 Le Mans, France}

\date{}

\maketitle

\begin{abstract}
We show that it is possible to replace the actual implicit distribution function of the
fractional exclusion statistics by an explicit one whose form does not change with the
parameter $\alpha$. This alternative simpler distribution function given by a generalization
of Pauli exclusion principle from the level of the maximal occupation number is not completely
equivalent to the distributions obtained from the level of state number counting of the
fractional exclusion particles. Our result shows that the two distributions are equivalent for
weakly bosonized fermions ($\alpha>>0$) at not very high temperatures.
\end{abstract}


\vspace{2cm}

The principle of the fractional exclusion statistics (FES) was for the first time proposed
about 60 years ago by Gentile\cite{Gentile} who suggested an intermediate maximum occupation
number changing from 1 (for fermions) to $\infty$ (for bosons). This idea was later recognized
and developed in the study of anyons and quasi-particle excitations for some low dimensional
systems relevant to fractional quantum Hall effect and to superconductivity\cite{Wilczek}.

Years ago, the study in this direction gets a new dimension by the consideration of the
influence of interactions on the number $d$ of one particle state\cite{Hald91,Wu01}. The
starting point is the assumption\cite{Hald91} that the variation of the number of states $d_i$
of a single particle energy level $i$ is proportional to the variation of the number of
particles $N_i$ occupying the level, i.e., $\Delta d_i=-\alpha \Delta N_i$ in a simple case
without mutual statistics\cite{Wu01}. Then by a state counting procedure\cite{Wu01} on the
basis of the assumption $d_i=G_i-\alpha(N_i-1)$ for ``bosons'' or $d_i=G_i-(1-\alpha)(N_i-1)$
for ``fermions'', one finds the number of quantum states $W$
($W_b=\frac{[d_i+(N_i-1)]!}{N_i![d_i-1]!}$ for ``bosons'' or
$W_f=\frac{[d_i]!}{N_i![d_i-N_i]!}$ for fermions) of $N_i$ particles distributed in the $G_i$
one particle states : $W_i=\frac{[G_i+(N_i-1)(1-\alpha)]!}{N_i![G_i-\alpha N_i-(1-\alpha)]!}$
interpolating between $W_b$ ($\alpha=0$) and $W_f$ ($\alpha=1$). Here $G_i$ can be considered
as the number of one particle states for ideal case. The total number of a system of $N$
particles and $\omega$ levels is then $W=\prod_{i=1}^\omega W_i$.

These generalized bosons or fermions can then be treated as ideal boson gas as
usual\cite{Wu01} under the additivity conditions of $E=\sum_i N_ie_i$ and $N=\sum_i N_i$ where
$E$ is the total energy of the system of $N$ particles and $e_i$ is the energy of one particle
level $i$. The most probable distribution is given by the average occupation number
$n_i=N_i/G_i$, i.e.
\begin{equation}                                                \label{1}
n_i=\frac{1}{f(e^{-\beta(e_i-\mu)})+\alpha}
\end{equation}
where $f(x)$ is a function satisfying
\begin{equation}                                                \label{1a}
f^{\alpha}(x)(1+f(x))^{1-\alpha} =x
\end{equation}
and $\alpha$ varies between $0\leq\alpha\leq 1$. The reader can find in Figure 1 this
distribution (full lines) for different $\alpha$ values. The constant $\alpha$ turns out to be
the inverse maximum occupation number ($1/\alpha$) of these fermionized bosons or bosonized
fermions. The exclusion principle of Pauli ($\alpha=0$ or 1) has been generalized here.

The functional form of Eq.(\ref{1}) depends on the value of $\alpha$. Eq.(\ref{1}) and
Eq.(\ref{1a}) do not necessarily have explicit solutions for any $\alpha$. In a previous
paper\cite{Wang03}, by avoiding the state counting procedure and using an usual method to
calculate the grand partition function, we proposed a simpler, explicit fractional
distribution which seemed identical to Eq.(\ref{1}) but whose functional form did not change
with $\alpha$.

The purpose of the present letter is to describe the method giving the alternative fractional
distribution and to provide a detailed analysis of the relationship between the two fractional
exclusion distributions. The reader will find that this approach does not need the two
different generalizations $d_i$'s for ``bosons'' and ``fermions''. It is only supposed that
the quasi-particles obey the fractional exclusion principle, i.e. the maximal number of
particles $n_m$ at a state may be different from 1 (fermions) or $\infty$ (bosons). In
average, $n_m<1$ may also make sense, so we propose $0<n_m<\infty$.

According to the hypothesis of the fractional ideal gas\cite{Wu01}, the $N$ quasi-particles
should be described by Boltzmann-Gibbs statistics. So that the grand partition function can be
given by\cite{Quarrie} (let $n_i=N_i$ for the moment) :
\begin{equation}                                            \label{2}
Z=\sum_{N=0}^\infty\sum_{\{n_i\}_N}e^{-\beta \sum_in_i(e_i-\mu)}
=\sum_{N=0}^\infty\sum_{\{n_i\}_N}\prod_i[e^{-\beta(e_i-\mu)}]^{n_i}
\end{equation}
where $\{n_i\}_N$ signifies all the possible sets $\{n_i\}$ obeying $N=\sum_in_i$ for a given
$N$. The summations in Eq.(\ref{2}) are equivalent to\cite{Quarrie}
\begin{equation}                                            \label{11}
Z=\prod_i\sum_{n_i=0}^{n_m}[e^{-\beta(e_i-\mu)}]^{n_i}
\end{equation}
The summation over $n_i$ is a geometric progression whose result is given by

\begin{equation}                                            \label{11a}
Z=\prod_i\frac{1-e^{-(1+n_m)\beta(e_i-\mu)}} {1-e^{-\beta(e_i-\mu)}}.
\end{equation}
Then as usual, the average occupation number $n$ of an one-particle state of energy $e$ is
calculated via the grand potential $\Omega=-kT\ln Z$ as follows
\begin{equation}                                 \label{12}
n=\frac{\partial\Omega}{\partial e} =-kT\frac{\partial(\ln Z)}{\partial e}
=\frac{1}{e^{\beta(e-\mu)}-1}- \frac{\frac{1+\alpha}{\alpha}}
{e^{\frac{1+\alpha}{\alpha}\beta(e-\mu)}-1}
\end{equation}
which recovers the standard boson distribution if $\alpha=0$ and the fermion distribution if
$\alpha=1$. Note that here we have put $n_m=1/\alpha$ according to the result of Eq.(\ref{1})
that $1/\alpha$ is the maximal occupation number.

Figure 1 presents a comparison of Eq.(\ref{12}) with Eq.(\ref{1}). Surprisingly, at low
temperature, Eq.(\ref{12}) (symbols) perfectly reproduces the distribution (full lines) given
by Eq.(\ref{1}). On the other hand, a small difference can be seen at higher temperatures.
This difference will be analyzed with the Fermi energies of the two distributions.

With Eq.(\ref{12}), it is straightforward to show that, if $T=0$, $n=1/\alpha$, $n=1/2\alpha$
and $n=0$ for $e<\mu$, $e=\mu$ and $e>\mu$, respectively. So the ``Fermi energy ($e_f^0$)" at
$T=0$ can be identified to $\mu$. For a {\it 2-dimensional gas}, $e_f^0=\alpha\frac{h^2N}{2\pi
mV}$ where $m$ is the mass of the particle and $V$ the volume of the system. When $T\neq 0$,
$e_f$ is determined by

\begin{equation}                                \label{13}
(e^{\frac{1+\alpha}{\alpha}\frac{e_f}{kT}}-1)/ (e^{\frac{e_f}{kT}}-1) = e^{\frac{e_f^0}{kT}}
\end{equation}
which is plotted (symbols) versus temperature in Figure 2 in comparison with the Fermi energy
(full lines) given by Eq.(\ref{1})\cite{Wu01}

\begin{equation}                                \label{13a}
e^{\frac{e_f}{kT}}=e^{\frac{e_f^0}{kT}}-e^{\frac{1-\alpha}{\alpha}\frac{e_f^0}{kT}}
\end{equation}
Note that in Figure 2 the particle density $N/V$ is chosen to give $e_f^0=1$ eV when
$\alpha=1$ (usual ideal fermions).

At low temperatures, there is no significant difference between the two Fermi-energies. But at
very high temperatures up to several K, important difference is observed for $\alpha$ very
different from 0 and unity. This difference is studied in Figure 3 through the relative
difference ($R$) between the two Fermi energies, where $R=\frac{\mid
e_f^{W}-e_f^{p}\mid}{\sqrt{\mid e_f^{W}e_f^{p}\mid}}$, $e_f^{W}$ and $e_f^{p}$ are the Fermi
energy given by Eq.(\ref{13a}) (lines) and that given by the present work in Eq.(\ref{13})
(symbols), respectively. For sufficiently large $\alpha$, $R$ is very small up to high
temperature. On the other hand, for small $\alpha$, $R$ may diverge at temperatures at which
$e_f^{W}$ or $e_f^{p}$ tends to zero.

In summery, a major result of this work is to provide an analysis of the relationship between
two possible distribution functions for fractional exclusion particles in the Haldane-Wu's
sense\cite{Hald91,Wu01}. The first was obtained by Wu\cite{Wu01} through state counting and
maximum entropy procedure on the basis of Haldane hypothesis $\Delta d=-\alpha \Delta
N$\cite{Hald91}. The second via the grand partition function calculated by supposing
$n_{max}=1/\alpha$ for the maximal occupation number. From the point of view of the fractional
Pauli exclusion principle, one would expect that these two approaches should be equivalent. In
fact, at temperatures up to ambiant ones, there is no significant difference between these two
distributions. However, for very small $\alpha$ (weakly fermionized bosons) or very high
temperatures, one should be careful because the relative difference between the two Fermi
energies may diverge at some temperatures. We hope that the explicit FES distribution function
of present work may be helpful for some calculations and applications.

It is worth noticing that the difference between the distributions Eq.(\ref{1}) and
Eq.(\ref{12}) means that {\it the generalization of Pauli exclusion principle at the level of
state number counting as described briefly at the beginning of this letter may be different
from that at the level of maximal occupation number} \`a la Gentile\cite{Gentile} as described
in this paper. This is a major conclusion of the present work.

\newpage

\begin{figure}[p] \label{f1}
\includegraphics[width=5in,height=4in]{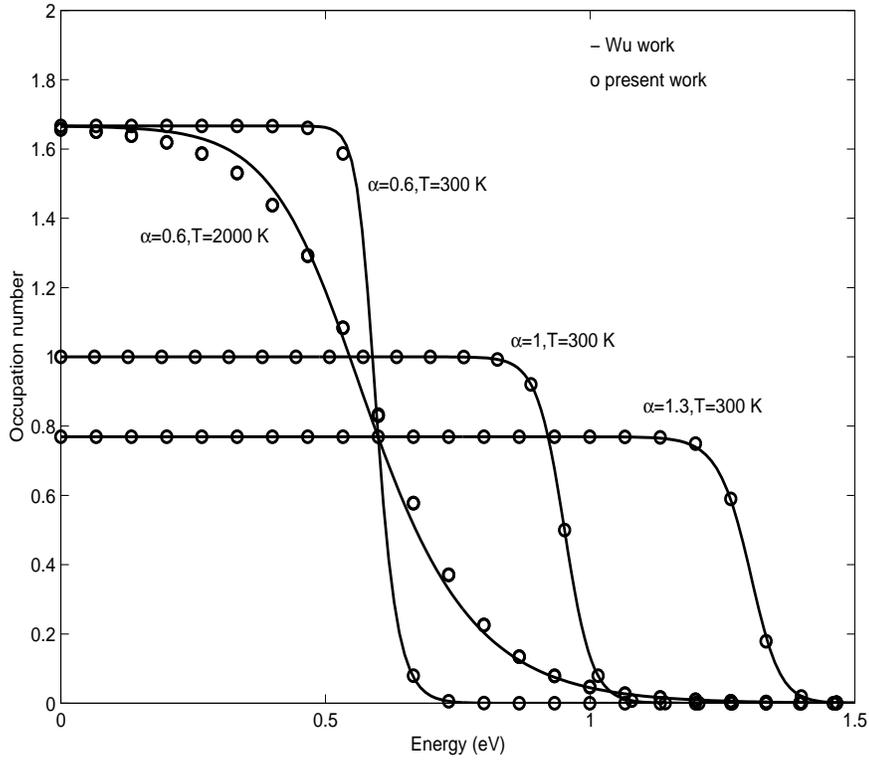}
\caption{Comparison of the FES distribution given in [4] by Wu [Eq.(\ref{1})] (lines) with
that given by the present work [Eq.(\ref{12})] (symbols) for different $\alpha$ values and
temperatures (300 K and 2000 K). We see that the two distributions are rather equivalent at
low temperature. A small difference takes place between these two distributions at high
temperature (see the lines for 2000 K).}
\end{figure}

\begin{figure}[p] \label{f2}
\includegraphics[width=5in,height=4in]{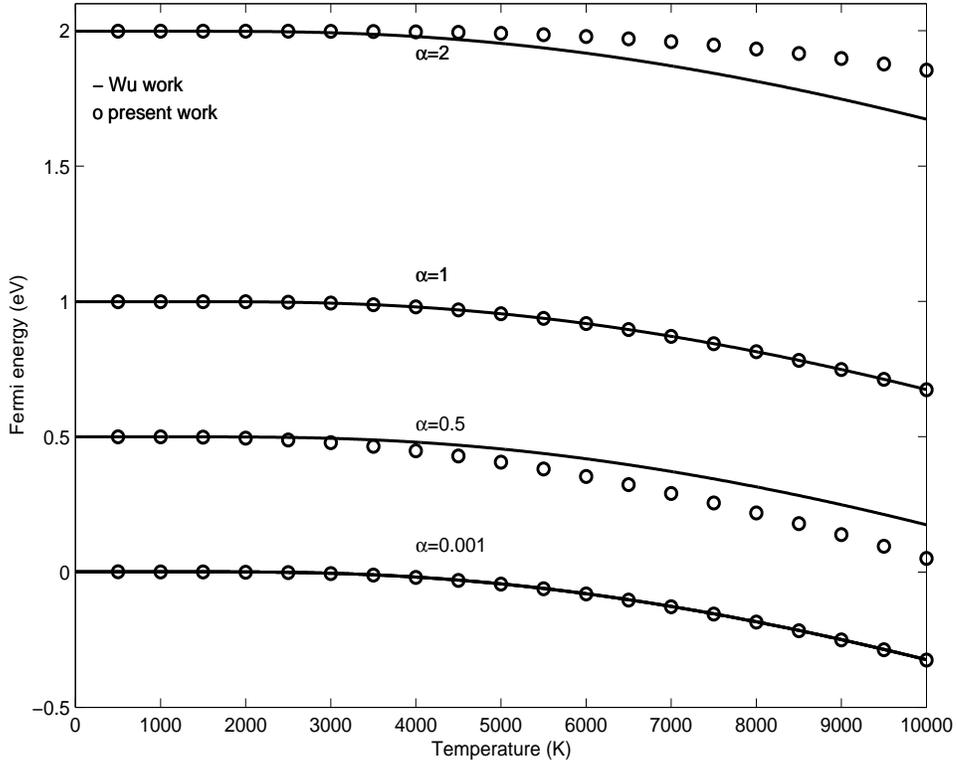}
\caption{Temperature dependence of 2-Dimensional Fermi energies given in [4] by Wu and by
present work for different values of $\alpha$. The are important differences at very high
temperatures for $\alpha$ values very different from 0 and unity. For low temperatures, the
two Fermi energies are sufficiently close each other. But for very small $\alpha$, the
relative difference may be important at low temperature as shown in Figure 3.}
\end{figure}

\begin{figure}[p] \label{f3}
\includegraphics[width=5in,height=4in]{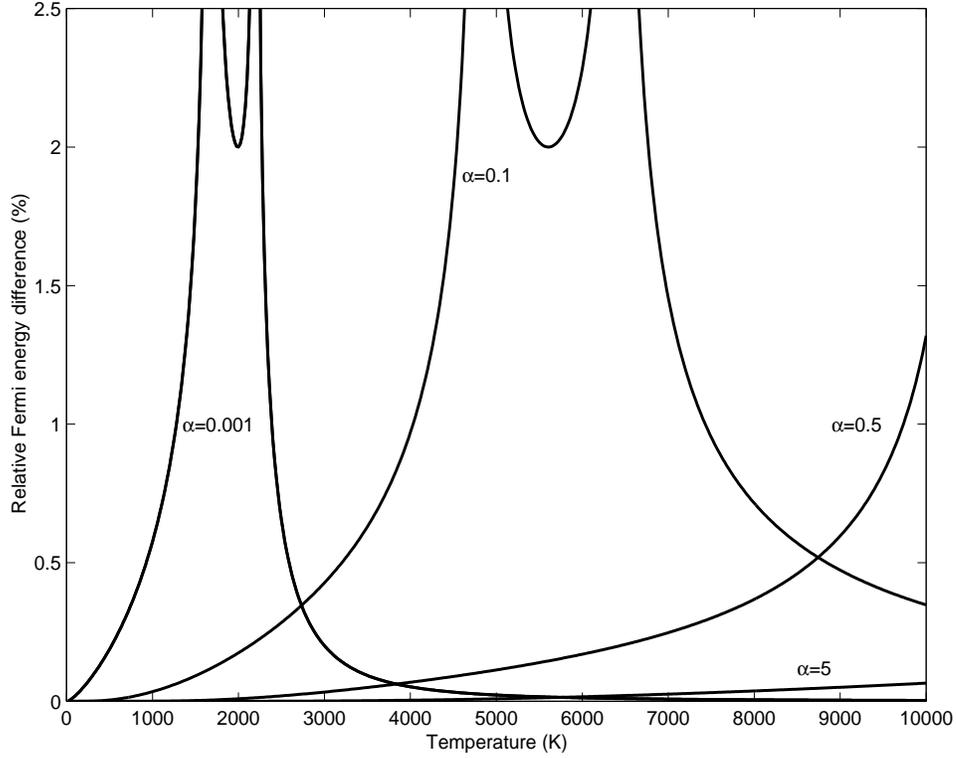}
\caption{Temperature dependence of the relative differences ($R$) between the two Fermi
energies plotted in Figure 2. Here $R=\frac{\mid e_f^{W}-e_f^{p}\mid}{\sqrt{\mid
e_f^{W}e_f^{p}\mid}}$, where $e_f^{W}$ and $e_f^{p}$ are the Fermi energy given by Wu in [4]
and that given by the present work, respectively. For sufficiently large $\alpha$, $R$ is very
small up to high temperature. On the other hand, for small $\alpha$, $R$ may diverge at
temperatures at which $e_f^{W}$ or $e_f^{p}$ tends to zero (the peaks).}
\end{figure}

\end{document}